\documentstyle[aps,prb,twocolumn,floats,psfig]{revtex}

\newcommand{\Y}{YNi$_2$B$_2$C~}
\newcommand{\Lu}{LuNi$_2$B$_2$C~}
\newcommand{\vq}{\bf q\rm}
\newcommand{\vk}{\bf k\rm}

\newcommand{\vH}{\bf H\rm}
\newcommand{\vv}{\bf v\rm}

\newcommand{\la}{\langle}
\newcommand{\ra}{\rangle}
\newcommand{\De}{$\Delta(\vk)~$}

\begin{document}

\draft

\title{Anisotropic s- wave superconductivity in borocarbides \Lu and \Y}

\author{K. Maki$^{1,2}$, P. Thalmeier$^3$ and H. Won$^4$}

\address{$^1$Max-Planck-Institute for the Physics of Complex Systems, 
N\"othnitzer Str.38, 01187 Dresden, Germany}
\address{$^2$Department of Physics and Astronomy,
University of Southern California, Los Angeles,}
\address{CA 90089-0484, USA}
\address{$^3$Max-Planck-Institute for the Chemical Physics of Solids,
N\"othnitzer Str.40, 01187 Dresden, Germany}
\address{$^4$Department of Physics, Hallym University, Chunchon 200-702,
South Korea}
\maketitle

\begin{abstract}
The symmetry of superconductivity in borocarbides \Lu and \Y is an
outstanding issue. Here an anisotropic s- wave order parameter (or s+g
model) is proposed for \Lu and \Y. In spite of the dominant s- wave
component the present superconducting order parameter \De has nodes
and gives rise to the
$\sqrt{H}$ dependent specific heat in the vortex state (the Volovik
effect). This model predicts the fourfold symmetry both in the angular
dependent thermal conductivity and in the excess Dingle temperature in
the vortex state, which should be readily accessible experimentally.
\end{abstract}


\section{Introduction}
The superconductivity in rare earth (R) transtion metal borocarbides
is of great interest\cite{Canfield98,Mueller01}, in particular its
interplay with magnetism and superconductivity is
fascinating\cite{Canfield98,Amici00}. However
in the following we limit ourselves to the nonmagnetic 
borocarbides \Lu and \Y. 
They have a relatively high superconducting transition
temperature T$_c$=15.5 K and 16.5 K respectively. Although the dominance
of the s-wave component in \De has been established by
substituting Ni by a small amount of Pt and subsequent opening of the
energy gap\cite{Nohara99,Borkowski94}, a
number of peculiarities are not expected in a conventional s- wave
superconductor\cite{Won01a}. For example the $\sqrt{H}$ dependence of
the specific heat in the vortex state indicates a superconducting
state with nodal excitations similar to d-wave superconductivity in high T$_c$
cuprates\cite{Volovik93,Nohara97,Freudenberger98,Izawa01a}.
Furthermore  the presence of de Haas van Alphen (dHvA) oscillations in
the vortex state of \Lu down to H=0.2H$_{c2}$ suggests again nodal
superconductivity\cite{Maki91,Wassermann93,Terashima97}. In a
conventional s-wave
superconductor dHvA oscillations would disappear for
H$<$0.8H$_{c2}$\cite{Maki91,Wassermann93}. In addition the upper critical
field determined for \Lu and \Y for field direction within the a-b
plane exhibits clear fourfold symmetry somewhat
reminiscent to d- wave superconductors\cite{Metlushko97,Wang98},
furthermore 1/T$_1$ from NMR experiments shows T$^3$ power law
behaviour consistent with nodal superconductors \cite{Zheng98}. These
experiments clearly indicate that \De in borocarbides has to be an anisotropic
s- wave order parameter. Furthermore a) \De has to have a
nodal structure with the quasiparticle density of states (DOS) N(E)$\sim|E|$
for $|E|$/$\Delta\ll$1, which gives the $\sqrt{H}$ dependence in the
specific heat of the vortex state\cite{Won01a,Volovik93}. b) the nodal
structure has to have the fourfold symmetry within the a-b plane which
is consistent with the tetragonal symmetry of the a-b plane. These two
constraints appear to suggest almost uniquely 

\begin{equation}
\label{GAP}
\Delta(\vk)=\frac{1}{2}\Delta(1+\sin^4\vartheta\cos(4\phi))
\end{equation}

or s+g- wave superconductivity. Here $\vartheta, \phi$ are the polar
and azimuthal angles in $\vk$- space respectively. We show in Fig. 1
\De which exhibits clear fourfold symmetry. The four second
order nodal points of \De are given by
$(\vartheta,\phi)$ = ($\frac{\pi}{2},\pm\frac{\pi}{4}$) and
($\frac{\pi}{2},\pm\frac{3\pi}{4}$) which dominate the quasiparticle
DOS at low energies:

\begin{equation}
\frac{N(E)}{N_0}=\frac{\pi}{4}\frac{|E|}{\Delta}+O(\frac{E}{\Delta})^2
\end{equation}

where N$_0$ is the normal state DOS. In
constructing \De, we have made use of a similar approach as in
MgB$_2$\cite{Nagamatsu01,Haas01,Chen01}. In the s+g model of
Eq.(\ref{GAP}) we assume the equality of s and g amplitudes to 
have N(E)$\sim |E|$ down to lowest energies. Recent thermal conductivity
measurements \cite{Boaknin01} report a gap anisotropy of at least a
factor of 10, the fine tuning of s and g amplitudes in Eq.(\ref{GAP}) therefore
has a tolerance of 10\%. There is no symmetry reason why the amplitudes
(or pair potentials) of inequivalent representations like s and g
should be very close. However from the bandstructure of borocarbides
\cite{Dugdale99} it may be argued that the pair potential at the nodal
points given above is indeed strongly suppressed. The main Fermi
surface sheet shows
lobe like structures along the [110] directions which have strong
nesting with a wave vector parallel to a. This wave vector appears 
as the incommensurate ordering vector in the magnetic borocarbides
(Lu,Y replaced by rare earth elements). Therefore the lobe states
at ($\vartheta,\phi$)=($\frac{\pi}{2},\pm\frac{\pi}{4}$) and
($\frac{\pi}{2},\pm\frac{3\pi}{4}$) tend to an instability in the
particle-hole channel which strongly depresses the effective potential
and associated \De for Cooper pairing at these points. The approximate
fine tuning (up to 10\%) of s and g amplitudes may be caused by this
peculiar Fermi surface feature of the borocarbides.

In the following we shall first consider thermodynamics and transport of the
borocarbides for zero field for the proposed gap function.
Then we will study the field angle dependence of specific
heat and thermal conductivity which exhibit the fourfold symmetry. We
apply the same technique  developed
in Ref.[\onlinecite{Volovik93,Won01b,Won01c,Thalmeier01}]. Also we
predict the fourfold symmetry in the excess Dingle temperature in the vortex
state in borocarbides in a planar magnetic field.

\begin{figure}
\centerline{\psfig{figure=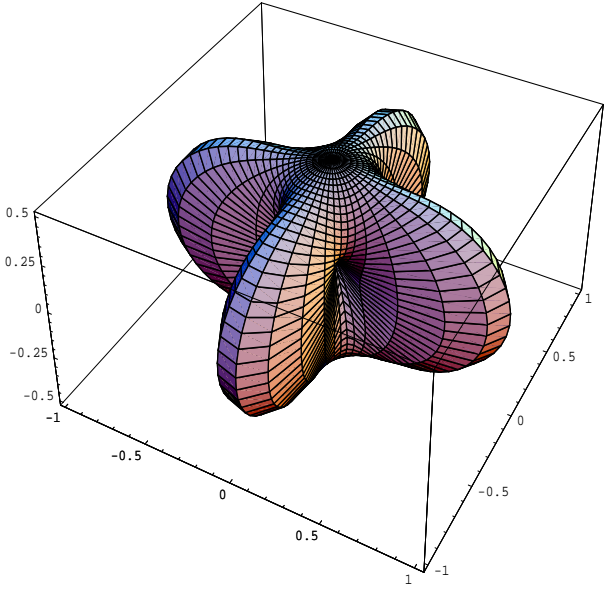,height=8cm,width=8cm}}
\vspace{1cm}
\caption{
Normalized gap function f(\vk)= \De/$\Delta$ of the s+g model. 
}
\end{figure}

\section{Thermodynamics and transport properties}

First of all \De =$\Delta$f(\vk) given in Eq.(\ref{GAP}) leads to the
quasiparticle density of states

\begin{eqnarray}
\frac{N(E)}{N_0}&=&\frac{1}{4\pi}\int d\Omega
Re\frac{|x|}{\sqrt{x^2-f^2}}\nonumber\\
&=&|x|\int_0^1dyF(y)Re\frac{1}{\sqrt{x^2-y^2}}
\end{eqnarray}

where x=E/$\Delta$ and 

\begin{eqnarray}
F(y)&=&\frac{2}{\pi}\int_0^{u_0}\frac{dz}
{\sqrt{(1-z^2)^4-(1-u_0^2)^4}}\nonumber\\
\mbox{with}~ u_0&=&(1-\sqrt{|1-2y|})^\frac{1}{2}
\end{eqnarray}

we note that F(1-y)=F(y) holds. The quasiparticle density of states is
evaluated numerically and shown in Fig.2. For $|E|$/$\Delta\ll$1 we obtain

\begin{figure}
\centerline{\psfig{figure=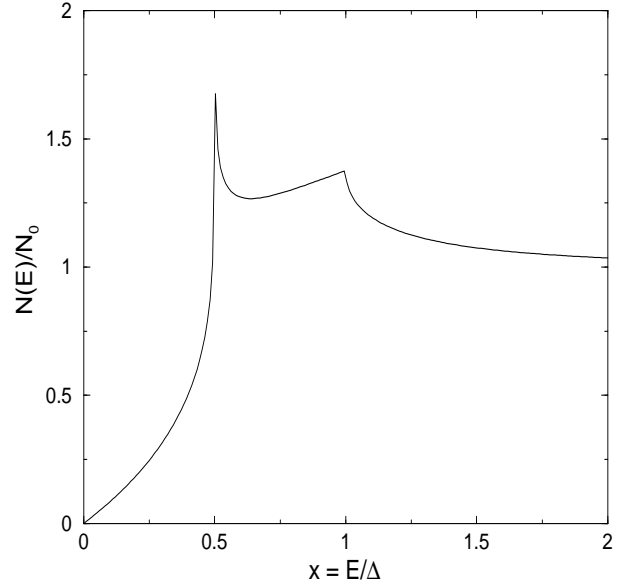,height=8cm,width=8cm}}
\vspace{1cm}
\caption{
Quasiparticle density of states. Logarithmic singularity occurs at
E=$\frac{\Delta}{2}$ due to the saddle points at
$\vartheta=0,\pi$. The cusp at E=$\Delta$ is due to the gap
maxima at ($\vartheta,\phi$)=($\frac{\pi}{2}$,0), ($\frac{\pi}{2},
\pm\frac{\pi}{2}$) and ($\frac{\pi}{2},\pi$). 
}
\end{figure}

\begin{eqnarray}
\frac{N(E)}{N_0}&=&\frac{\pi}{4}\frac{|E|}{\Delta}
(1+\frac{9}{4\pi}\frac{|E|}{\Delta}+...)
\end{eqnarray}

then the low temperature specific heat is given by

\begin{eqnarray}
\frac{C_s}{\gamma_NT}=\frac{27}{4\pi}\zeta(3)(\frac{T}{\Delta})
+\frac{63}{80}(\frac{\pi T}{\Delta})^2+..
\end{eqnarray}

where $\gamma_N$ is the Sommerfeld constant. Similarly the spin
susceptibility and the superfluid density are given by\cite{*}

\begin{eqnarray}
\frac{\chi}{\chi_N}&=&\frac{\pi}{2}\frac{T}{\Delta}(\ln 2)
+\frac{3\pi^2}{16}(\frac{T}{\Delta})^2+..\nonumber\\
\frac{\rho_s(T)}{\rho_s(0)}&=&1-\frac{\chi}{\chi_N}
\end{eqnarray}

Likewise the electronic thermal conductivity of the s+g model at low
temperature is obtained in a universal form 
as 

\begin{eqnarray}
\frac{\kappa}{T}=\frac{\pi^2}{8}\frac{n}{m\Delta}
\end{eqnarray}

The prefactor $\pi^2$/8 is specific for the s+g model. Here n, m are
the electronic density and mass respectively.
This is equivalent to $\kappa/\kappa_n$ =$3\Gamma/8\Delta$ where
$\kappa_n$ is the normal state thermal conductivity and $\Gamma$ the
quasiparticle scattering rate. The linear T behaviour of $\kappa$ has
recently been found \cite{Boaknin01} in \Lu from which we extract
$\Gamma/\Delta\leq$ 0.02.

\section{Angular dependent specific heat and thermal conductivity}

We are proposing that the angular dependent specific heat and
especially thermal conductivity in
the vortex state provides a unique window to look for the symmetry of
\De \cite{Won01b,Won01c,Thalmeier01,Won01d}. Indeed from the latter
Izawa et al have succeded in deducing the symmetry of \De
in Sr$_2$RuO$_4$\cite{Izawa01b}, CeCoIn$_5$\cite{Izawa01c} and more recently
$\kappa$-(ET)$_2$Cu(NCS)$_2$\cite{Izawa01d}. First of all we have to
construct the equation for the residual density of states in the
presence of impurity scattering\cite{Barash97}. 

\begin{eqnarray}
\label{GFUNC}
g&=&\it Re\rm\Biggl\la\frac{C_0-ix}
{\sqrt{(C_0-ix)^2+f^2}}\Biggr\ra \nonumber\\
&&\frac{1}{2}\sum_\pm\Bigl\la C_0\ln(\frac{2}{\sqrt{C_0^2+x^2}})+
x\tan^{-1}(\frac{x}{C_0})\Bigr\ra
\end{eqnarray}

where C$_0$=$\lim_{\omega\rightarrow 0}Im(\tilde{\omega}/\Delta$) with 
$\tilde{\omega}$ giving the renormalized frequency and
x= $|\vv\cdot\vq|$/$\Delta$ $\sim|\sin(\theta\pm\frac{\pi}{4})|$. 
Here 2$\vq$ is the sum of the pair momentum associated with a
supercurrent circulating around each vortex and $\vv\cdot\vq$ is the
Doppler shift connected with it. In the first line the brackets mean
averaging over both Fermi surface and vortex lattice, in the second
line the former is evaluated up to the $\pm$ summation and the latter
still remains. In the superclean limit defined by C$_0\ll\la x\ra$ or
$\Gamma\ll v_a\sqrt{eH}\ll T\ll\Delta$ Eq.(\ref{GFUNC}) gives

\begin{eqnarray}
g=\frac{\pi}{4}\la x\ra = \frac{\tilde{v}\sqrt{eH}}{2\sqrt{2}\Delta}I(\theta)
\end{eqnarray}

where $\tilde{v}=\sqrt{v_av_c}$ and I($\theta$)=$\max{(|\sin
\theta|,|\cos \theta|)}$ for 0$\leq\theta\leq\frac{\pi}{2}$. The
function I($\theta$) is shown in Fig. 3. Here v$_a$ and v$_c$ are Fermi
velocities in the a-b plane and along the c-axis respectively. The
magnetic field is applied within the a-b plane at an angle 
$\theta$ with respect to the a axis.

In the clean limit with C$_0\gg\la x\ra$ or $ v_a\sqrt{eH}\ll\Gamma\ll
T\ll\Delta$ on the other hand we obtain 

\begin{eqnarray}
g=g(0)\Bigl(1+\frac{\tilde{v}^2(eH)}{32\Gamma\Delta}
\bigl[\ln(\frac{\Delta}{\tilde{v}\sqrt{eH}}
-\frac{1}{8}(1-\cos(4\theta))\bigr]\Bigr)
\end{eqnarray}

From these expressions the field angular dependent specific heat in
the vortex state may be derived. In the superclean limit we obtain
 
\begin{eqnarray}
\frac{C_s}{\gamma_NT}=\frac{\tilde{v}\sqrt{eH}}{\sqrt{2}\Delta}I(\theta)
\end{eqnarray}

In the clean limit on the other hand the above $\theta$ dependence is
replaced by 

\begin{eqnarray}
\frac{C_s}{\gamma_NT}&=&g(0)\Bigl(1+\frac{\tilde{v}^2(eH)}{32\Gamma\Delta}
\bigl[\ln(\frac{\Delta}{\tilde{v}\sqrt{eH}})
-\frac{1}{8}(1-cos(4\theta))\bigr]\Bigr)\nonumber\\
&&
\end{eqnarray}

where g(0)=N(0)/N$_0$ in the absence of magnetic field. 

The thermal conductivity tensor in the vortex phase has been
calculated in \cite{Won01c} and in a planar magnetic field it is given
by 

\begin{eqnarray}
\frac{\kappa_{xx}}{\kappa_n}&=&
\frac{3}{32}\frac{\tilde{v}^2(eH)}{\Delta^2}I^2(\theta)\nonumber\\
\frac{\kappa_{xy}}{\kappa_n}&=&
-\frac{3}{64}\frac{\tilde{v}^2(eH)}{\Delta}\sin(2\theta)\nonumber\\
\end{eqnarray}

in the superclean limit and

\begin{eqnarray}
\frac{\kappa_{xx}}{\kappa_0}&=&
1+\frac{\tilde{v}^2(eH)}{12\Gamma\Delta}
\ln(2\sqrt{\frac{\Delta}{\Gamma}})[\ln(\frac{\Delta}{\tilde{v}\sqrt{eH}})
\nonumber\\
&&-\frac{1}{8}(1-\cos(4\theta))] \\
\frac{\kappa_{xy}}{\kappa_0}&=&
-\frac{\tilde{v}^2(eH)}{12\Gamma\Delta}\sin(2\theta)
\ln(2\sqrt{\frac{\Delta}{\Gamma}})\ln(\frac{\Delta}{\tilde{v}\sqrt{eH}})
\nonumber
\end{eqnarray}

in the clean limit. Here $\kappa_0$ is $\kappa_{xx}$(H=0). Therefore
we expect the fourfold symmetry in the thermal conductivity in the
vortex state should be readily accessible in future experiments. 
On the other hand $\kappa_{xx}$ has recently been measured for field
oriented along c \cite{Boaknin01}. In this case a similar calculation
in the superclean limit for H$\ll$ H$_{c2}$ leads to 

\begin{eqnarray}
\frac{\kappa_{xx}(H)}{\kappa_n}=\frac{3}{\pi}
\frac{v_a^2(eH)}{\Delta^2(0)}\sim\frac{H-H_{c1}}{H_{c2}(0)}
\end{eqnarray}

This behaviour was indeed experimentally observed in \cite{Boaknin01}.
In the clean limit $\kappa_{xx}$(H) is no longer exactly linear but
has a logarithmic correction in H. Since $\Gamma/\Delta\leq$0.02 for
\Lu we can use the above equation for the superclean limit except for
very small fields.

\section{Excess Dingle temperature}

It is well known that dHvA oscillations can be seen in the vortex
state as well when the quasiparticle damping is much less than the
cyclotron frequency\cite{Maki91,Maki93,Corcoran95}. However in
conventional s-wave superconductors the dHvA oscillation becomes
invisible when H$\leq$0.8
H$_{c2}$. Therefore if dHvA oscillations are seen even for
H$\sim$0.2H$_{c2}$ as in the case of \Lu\cite{Terashima97}, this can
be taken as a signature of a nodal superconductor. Since the excess Dingle
temperature in the vortex state is due to quasiparticle damping caused
by the Andreev
scattering it should also exhibit the fourfold symmetry of the order
parameter. The excess damping due to Andreev scattering is evaluated
as 

\begin{eqnarray}
\Gamma_A&=&\frac{\pi}{2\tilde{v}}\frac{1}{\sqrt{eH}}\la\Delta^2\ra \nonumber\\
&&=\frac{\pi}{2}\frac{1}{\tilde{v}\sqrt{eH}}\Delta^2J(\theta)
\end{eqnarray}

where we defined

\begin{eqnarray}
J(\theta)&=&\frac{1}{4\pi}\int_{-\frac{\pi}{2}}^{\frac{\pi}{2}}
d\vartheta(1+\sin^4\vartheta\cos(4\theta))^2 \nonumber\\
&&=\frac{1}{4}(1+\frac{3}{4}\cos(4\theta)+\frac{35}{128}\cos^2(4\theta))
\end{eqnarray}

\begin{figure}
\centerline{\psfig{figure=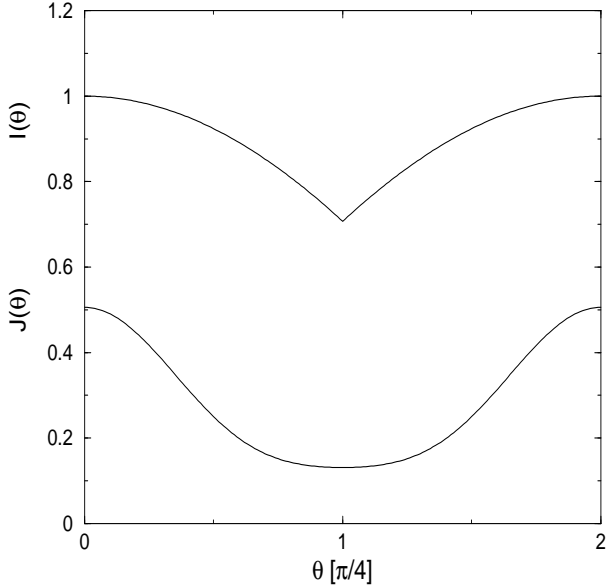,height=8cm,width=8cm}}
\vspace{1cm}
\caption{
Angular dependence of specific heat C$_s\sim$ I($\theta$) and excess
Dingle temperature $\sim$ J($\theta$) in an external field in the a-b
plane. $\theta$ is the angle between field direction and a- axis.
}
\end{figure}

That is we average $\Delta^2(\vk)$ on the Fermi surface sliced
perpendicular to $\vH$.
The angular dependence of J($\theta$) is shown in Fig. 3 together with
I($\theta$). In particular we find I($\frac{\pi}{4}$)/I(0)=
$\frac{1}{\sqrt{2}}$ and J($\frac{\pi}{4}$)/J(0)= 0.2587. The
excess damping is reduced by a factor of $\frac{1}{4}$ for
$\vH\parallel$ [1,1,0] as compared to the one for $\vH\parallel$ [1,0,0]
for example. 

\section{Concluding Remarks}

Here we propose a simple model for \De for nonmagnetic borocarbide
superconductors with fourfold symmetry. The angular dependence of the
specific heat, thermal conductivity and the excess Dingle temperature
are worked out with this model. We hope that this work will stimulate
further experiments on borocarbide superconductors.\\

\noindent
{\em Acknowledgement}\\ 
We would like to thank Koichi Izawa and Yuji Matsuda for useful
discussions on superconducting borocarbides. K.M. also thanks the
hospitality of Department of Physics at Hallym University where a part
of this work was done. H.W. acknowledges the support of the KOSEF 
through CSCMR.

\end{document}